\documentclass[12pt]{article}

\renewcommand{\d}{{\rm d}}
\newcommand{\gsth}{g_{\sigma^3}}
\newcommand{\dsth}{\delta_{\sigma^3}}
\newcommand{\sgn}{{\rm sgn}}
\newcommand{\Dsth}{\Delta_{\sigma^3}}
\newcommand{\sth}{\sigma^3}
\newcommand{\Msth}{M_{\sigma^3}}
\newcommand{\tilMsth}{\tilde{M}_{\sigma^3}}
\newcommand{\Vsth}{V_{\sigma^3}}
\newcommand{\bx}{\mbox{\boldmath$x$}}

\newcommand{\sigd}{\sigma^d}
\newcommand{\Dsigd}{\Delta_{\sigma^d}}
\newcommand{\Vsigd}{V_{\sigma^d}}
\newcommand{\ltwosigd}{l^2_{\sigma^d}}
\newcommand{\ltwosth}{l^2_{\sigma^3}}
\newcommand{\stw}{\sigma^2}
\newcommand{\son}{\sigma^1}
\newcommand{\sfo}{\sigma^4}
\newcommand{\ltwoson}{l^2_{\sigma^1}}
\newcommand{\Dsthk}{\Delta_{\sigma^3_k}}
\newcommand{\ltwosonsfok}{l^2_{\sigma^1\sigma^4_k}}

\newcommand{\ltwosonsfokplon}{l^2_{\sigma^1\sigma^4_{k+1}}}
\newcommand{\dthstwnul}{\delta^3_{\sigma^2}(0)}
\newcommand{\dsonnul}{\delta_{\sigma^1}(0)}
\newcommand{\Dstw}{\Delta_{\sigma^2}}
\newcommand{\gstw}{g_{\sigma^2}}
\newcommand{\dfun}{$\delta$-function }
\newcommand{\dfuns}{$\delta$-functions }
\newcommand{\dsigd}{\delta_{\sigma^d}}
\newcommand{\Vstw}{V_{\sigma^2}}
\newcommand{\Vson}{V_{\sigma^1}}
\newcommand{\Sind}{S_{\rm indep}}
\newcommand{\dmuind}{\d \mu_{\rm indep}}
\newcommand{\sumstw}{\sum_{\sigma^2}}
\newcommand{\sumsfosupstw}{\sum_{\sigma^4 \supset \sigma^2}}
\newcommand{\alstwsfo}{\alpha_{\sigma^2\sigma^4}}
\newcommand{\Astw}{A_{\sigma^2}}
\newcommand{\Astwsfo}{A_{\sigma^2\sigma^4}}
\newcommand{\sumsfo}{\sum_{\sigma^4}}
\newcommand{\Nstw}{N_{\sigma^2}}
\newcommand{\sumstwsubsfo}{\sum_{\sigma^2 \subset \sigma^4}}

\begin{document}

\title{Gravity action on the rapidly varying metrics}

\author{V.M. Khatsymovsky \\
 {\em Budker Institute of Nuclear Physics} \\ {\em
 Novosibirsk,
 630090,
 Russia}
\\ {\em E-mail address: khatsym@inp.nsk.su}}
\date{}
\maketitle

\begin{abstract}
We consider a four-dimensional simplicial complex and the minisuperspace general relativity system described by the metric flat in the most part of the interior of every 4-simplex with exception of a thin layer of thickness $\propto \varepsilon$ along the every three-dimensional face where the metric undergoes jump between the two 4-simplices sharing this face. At $\varepsilon \to 0$ this jump would become discontinuity. Since, however, discontinuity of the (induced on the face) metric is not allowed in general relativity, the terms in the Einstein action tending to infinity at $\varepsilon \to 0$ arise. In the path integral approach, these terms lead to the pre-exponent factor with \dfuns requiring that the induced on the faces metric be continuous, i. e. the 4-simplices fit on their common faces. The other part of the path integral measure corresponds to the action being the sum of independent terms over the 4-simplices. Therefore this part of the path integral measure is the product of independent measures over the 4-simplices. The result obtained is in accordance with our previous one obtained from the symmetry considerations.

\end{abstract}

PACS numbers: 31.15.xk; 04.60.Kz

MSC classes: 83C27; 81S40

\section{Introduction}
\label{intro}

Simplicial general relativity or Regge calculus \cite{Regge} is minisuperspace gravity theory. This means that it is genuine general relativity although formulated for the particular set of Riemannian metrics, the so-called piecewise-flat ones. The latter are flat everywhere with exception of the set of zero measure. Any such manifold can be viewed as collection of the flat 4-dimensional tetrahedra or 4-simplices $\sigma^4$. The curvature distribution residues on the 2-faces, i. e. triangles or 2-simplices $\sigma^2$. The curvature develops as nonzero angle defect which is deviation of the sum of hyperdihedral angles meeting at the 2-simplex $\sigma^2$ from $2\pi$. In the paper \cite{Fri} the Regge action \cite{Regge} is obtained by direct substitution of the piecewise-flat metric into the Einstein action. Suppose we use affine metric in each flat 4-simplex. We can observe that generally the component of such metric along the normal to a 3-dimensional face $\sigma^3$ separating a pair of the 4-simplices $\sigma^4_1$, $\sigma^4_2$ suffers from discontinuity when passing from $\sigma^4_1$ to $\sigma^4_2$. This is necessary condition for the angle defect to appear. At the same time, tangential or induced on $\sigma^3$ metric is still unambiguous if defined either in $\sigma^4_1$ or in $\sigma^4_2$ --- this is simply condition that 4-tetrahedrons $\sigma^4_1$ and $\sigma^4_2$ fit on their common 3-face $\sigma^3$. That is, the lengths of the coinciding edges of $\sigma^4_1$ and $\sigma^4_2$ are the same.

The Einstein action is of course not defined if the induced on the 3-faces metric is discontinuous: general relativity adopts only physical configurations. Such metric would correspond to the case when the different 4-simplices are allowed to have not the same lengths of their common edges. However, we may consider the case when the jumps of the induced on the 3-faces metric are not quite sharp but are smoothed over the thin intermediate layers of the thickness $\propto \varepsilon$ in the neighborhood of the 3-faces. Then we get infinities in the Einstein action at $\varepsilon \to 0$ which make it impossible for the discontinuities of the induced on the 3-faces metric to exist on classical level. In the path integral formalism these infinities turn out to result in the \dfun pre-exponent factor requiring the considered discontinuities to vanish. The form of this \dfun factor is just the subject of the analysis of the present paper.

Exact form of the considered pre-exponent factor depends on the details of tending to zero the thickness of the intermediate layers on the different 3-faces as $\varepsilon \to 0$. This point and arguments in favor of the chosen way of performing this limit are discussed in the Conclusion. As for the form of the metric jump across the 3-faces, it is taken rather general. The path integral measure has the form of the \dfun containing pre-exponent factor times the other part of the measure corresponding to an action being the sum of the independent terms over the 4-simplices. Formally, the latter action corresponds to the dynamics of the independent 4-simplices, and the measure looks as a result of some projection of the measure in the extended configuration superspace of independent 4-simplices onto the physical hypersurface of the simplicial general relativity. The problem of such projection of the measure from the superspace of independent 4-simplices was considered in our paper \cite{Kha} from the viewpoint of symmetry considerations. (In the present derivation we do not consider the theories with nonzero metric discontinuities \cite{WaiWil}.) The result of the present paper for the path integral measure coincides with that of our paper \cite{Kha}.

\section{Simple example}\label{1dim-example}

Consider a system with $N$ degrees of freedom described by the coordinates $x^A(t), A = 1, 2, \dots , N$ with the action
\begin{equation}
S = \frac{1}{2} \int M_{AB} (x) \dot{x}^A \dot{x}^B \d t , ~~~ M_{AB} = M_{BA}.
\end{equation}

\noindent Canonical Hamiltonian path integral for this system reads
\begin{eqnarray}\label{dpdq}
& & \int \exp \left \{ i \int \left [ p_A \dot{x}^A - \frac{1}{2} (M^{-1})^{AB} p_A p_B \right ] \d t \right \} \prod_{t, A} \d p_A (t) \d x^A (t) \nonumber \\ & & \propto \int \exp \left [ \frac{i}{2} \int M_{AB} (x) \dot{x}^A \dot{x}^B \d t \right ] \prod_t \left \{ [ \det M (x(t)) ]^{1/2} \prod_A \d x^A (t) \right \}.
\end{eqnarray}

Suppose there is jump of $x^A (t)$,
\begin{equation}
x^A (t) = f (t / \varepsilon) \Delta x^A + x^A_1 , ~~~ \Delta x^A = x^A_2 - x^A_1 , ~~~ f(- 1 / 2 ) = 0, ~~~ f( + 1 / 2 ) = 1.
\end{equation}

\noindent The action is the sum of the divergent at $\varepsilon \to 0$ term $O ( 1 / \varepsilon )$ and of the finite part $\Sind = O(1)$ being the sum of the actions for the two independent pieces of the trajectory,
\begin{eqnarray}\label{S-indep}
& & S = \frac{1}{2} \frac{\Delta x^A \Delta x^B}{\varepsilon } \int^{+1/2}_{-1/2} M_{AB} [f(\tau) \Delta x + x_1 ] f^{\prime 2} (\tau ) \d \tau + \Sind , \nonumber \\ & & \Sind = \frac{1}{2} \int_{t \not \in [-\varepsilon / 2 , + \varepsilon / 2 ]} M_{AB} (x) \dot{x}^A \dot{x}^B \d t
\end{eqnarray}

\noindent The divergent term results in the \dfun of $\Delta x$ as a factor in the path integral since
\begin{equation}\label{delta-x}
\exp (ix^2/\varepsilon ) = \sqrt{\pi\varepsilon} \exp (i\pi/4) \delta (x)
\end{equation}

\noindent at $\varepsilon \to 0$ as distribution on a standard set of probe functions. The overall pre-exponent factor reads
\begin{eqnarray}
& & \exp \left \{ \frac{i}{2} \frac{\Delta x^A \Delta x^B}{\varepsilon } \int^{+1/2}_{-1/2} M_{AB} [f(\tau) \Delta x + x_1 ] f^{\prime 2} (\tau ) \d \tau \right \} \nonumber \\ & & = \frac{(2 \pi \varepsilon )^{N / 2} \exp (i \pi N / 4) \delta^N (\Delta x) }{\sqrt{\det \left \{ \int^{+1/2}_{-1/2} M_{AB} ( x_1 ) f^{\prime 2} (\tau ) \d \tau \right \}}} \nonumber \\ & & = (2 \pi \varepsilon )^{N / 2} \exp (i \pi N / 4) \left [ \int^{+1/2}_{-1/2} f^{\prime 2} (\tau ) \d \tau \right ]^{-1/2} [\det M(x_1)]^{-1/2} \delta^N (\Delta x).
\end{eqnarray}

\noindent Here $\arg \det M = N_- \pi$, $N_-$ being the number of the negative eigenvalues of the matrix $M$. It is natural that namely dimensionless function of the dimensionless parameter $f(\tau )$ be universal one for every jump of the trajectory $x^A (t)$. Then $\int^{+1/2}_{-1/2} f^{\prime 2} (\tau ) \d \tau$ is some universal constant. Up to a constant factor, the path integral reads
\begin{eqnarray}
& & \exp \left [ \frac{i}{2} \int^{-0} M_{AB} (x) \dot{x}^A \dot{x}^B \d t + \frac{i}{2} \int_{+0} M_{AB} (x) \dot{x}^A \dot{x}^B \d t \right ] \nonumber \\ & & \cdot \{ \det M [ x ( -0 ) ] \}^{-1/2} \delta^N [ x ( +0 ) - x ( -0 ) ] \prod_t \{ \det M [ x ( t ) ] \}^{1/2} \d^N x(t).
\end{eqnarray}

\noindent The arising \dfun removes discontinuity of the path a priori possible at $\varepsilon \to 0$. Subsequent integration over $\d^N x(+0)$ will remove the \dfun itself, and we shall return to the original form of the measure (\ref{dpdq}). In fact, this is quite natural to expect a priori. Indeed, we have singled out the trajectories with arbitrarily sharp jumps, but the region of integration in the path integral already contains such trajectories. Therefore we have obtained something like scalability property for the measure: the measures for the independent pieces of a trajectory are glued into the measure for the overall trajectory via introducing the arbitrarily sharp jumps between the pieces similar to the jumps which every piece can already contain.

Vice versa, requiring that inclusion of the paths with arbitrarily sharp jumps would not change the form of the measure we might get the integration element in the measure $\prod_t \{ \det M [ x ( t ) ] \}^{1/2} \d^N x(t)$. In this sense the considered procedure of inclusion of these paths is consistent with the canonical Hamiltonian approach which gives the same integration element in the measure.

Thus, extending the region of integration by inclusion of the paths with arbitrarily sharp jumps gives for the path integral measure
\begin{equation}\label{dmu-dmu-indep}
\d \mu = \dmuind \prod_i \{ \det M [ x ( t_i -0 ) ] \}^{-1/2} \delta^N [ x ( t_i +0 ) - x ( t_i -0 ) ].
\end{equation}

\noindent Here $\dmuind$ is a path integral measure for the action $\Sind$ which is the sum of the actions on the parts of the path between the jumps. Since these parts are independent, the $\dmuind$ should naturally be the product of the independent path integral measures (on the intervals of time axis). In the particular case when the measure on each part of the path is canonically derived from the Hamiltonian approach, the resulting $\d \mu$ is canonical one in the same sense too.

\section{The case of gravity}

Consider analog of the $O(1 / \varepsilon)$ term in Eq. (\ref{S-indep}) for the case of the general relativity action. We use the metric $g_{\lambda\mu}$ with the signature $(-,+,+,+)$, $\lambda$, $\mu$, \dots = 0,1,2,3. It is assumed constant in the every 4-simplex $\sigma^4$ with exception of a thin intermediate layer in the vicinity of each its 3-face. Let $x^n$ be coordinate normal to the considered 3-face $\sigma^3$. The metric is chosen as $\d s^2 = g_{nn}(\d x^n)^2 + g_{\alpha\beta}\d x^{\alpha} \d x^{\beta}$. Here the indices $\alpha$, $\beta$, \dots take on some 3 values of 0, 1, 2, 3; $n$ is the 4th value. Denote the matrix $\| g_{\alpha\beta} \|$ for the given $\sigma^3$ as $\gsth$. The 3-face $\sigma^3$ can be timelike or spacelike one, $\sgn \det \gsth$ = $- \sgn g_{nn}$. Suppose there is a jump of $\gsth$ across $\sth$,
\begin{equation}
\gsth (x^n) = f (x^n / \Delta x^n ) \Delta \gsth + \gsth^{(1)}, ~~ \Delta \gsth = \gsth^{(2)} - \gsth^{(1)}, ~~ f(- 1 / 2 ) = 0, ~~ f( + 1 / 2 ) = 1,
\end{equation}

\noindent between the 4-simplices $\sfo_1$ ($x^n < 0$) and $\sfo_2$ ($x^n > 0$). The $\gsth^{(1)}$, $\gsth^{(2)}$ are the values of the matrix $\gsth$ inside these 4-simplices outside the intermediate layer. The singular at $\varepsilon \to 0$ term in the gravity action $S_{\rm g}$ contributed by the neighborhood of the given 3-face $\sth$ is
\begin{equation}\label{S-gsing}
S_{\rm g \, sing} = -{1\over 64 \pi G} \int (g^{\alpha\gamma}g^{\beta\delta} - g^{\alpha\beta}g^{\gamma\delta})g_{\alpha\beta , n}g_{\gamma\delta , n}g^{nn}\sqrt{-\det g}\d^4x.
\end{equation}

Locally in the neighborhood of a 3-face the system looks like the one-dimensional one of the Section \ref{1dim-example}. The main difference is that the system is in fact discrete and is lacking in the set of variables $q(t)$ parameterized by the continuous parameter $t$ (time). Therefore the canonical Hamiltonian formalism is not defined. However, the structure and origin of the \dfun pre-exponent factor remains the same.

Denote the matrix of the bilinear in $g_{\alpha\beta ,n}$ form in (\ref{S-gsing}) (the function of $\gsth$) by
\begin{equation}\label{M}
M^{(\alpha\beta)(\gamma\delta)}_{\sth} \equiv M^{(\alpha\beta)(\gamma\delta)} (\gsth ) = {1\over 2}(g^{\alpha\gamma}g^{\beta\delta} + g^{\alpha\delta}g^{\beta\gamma}) - g^{\alpha\beta}g^{\gamma\delta}.
\end{equation}

\noindent Also we can adopt $g_{nn}$ to vary between the 4-simplices,
\begin{equation}
g_{nn} (x^n) = \theta ( x^n / \Delta x^n ) \Delta g_{nn} + g^{(1)}_{nn} , ~~~ \theta ( - 1 / 2 ) = 0, ~~~ \theta ( + 1 / 2 ) = 1.
\end{equation}

\noindent Then we get the following bilinear form over $\Delta \gsth$,
\begin{equation}
S_{\rm g \, sing} = - \frac{\Delta^3 \bx }{64 \pi G \Delta x^n} (\Delta \gsth \tilMsth \Delta \gsth ) ,
\end{equation}

\noindent with the matrix
\begin{eqnarray}
\tilMsth = \int^{+1/2}_{-1/2} \sqrt{\frac{- \det \left [ f(\tau ) \Delta \gsth +\gsth^{(1)}\right ]}{\theta (\tau ) \Delta g_{nn} + g^{(1)}_{nn}}} M \left [ f(\tau ) \Delta \gsth +\gsth^{(1)}\right ] f^{\prime 2} (\tau ) \d \tau .
\end{eqnarray}

\noindent Here $\sqrt{\det \gsth } \Delta^3 \bx / 6 = \Vsth$ is the volume of the 3-face $\sth$ (real for the spacelike $\sth$). The $\exp (i S_{\rm g \, sing})$ at $\Delta x^n \to 0$ gives the \dfun $\delta^6 (\Delta \gsth)$ which enforces $\Delta \gsth = 0$ in $\tilMsth$. It is natural to take $f(\tau )$ symmetric w. r. t. the both 4-simplices, $f(\tau ) = 1 - f(-\tau )$. The $g_{nn}$ generally can be discontinuous, and the Heaviside step function can be taken for $\theta (\tau )$. We denote
\begin{equation}
\overline{g^{-1/2}_{nn}} \equiv \frac{1}{2} [g^{(1)}_{nn}]^{-1/2} + \frac{1}{2} [g^{(2)}_{nn}]^{-1/2}
\end{equation}

\noindent and collect the factors into the following dimensionless infinitesimal parameter $\varepsilon$ which will be used instead of $\Delta x^n$,
\begin{equation}\label{varepsilon}
\frac{1}{\varepsilon} = - \frac{\Delta^3 \bx \left [ - \det \gsth^{(1)}\right ]^{1 / 2} }{64 \pi G \Delta x^n} \overline{g^{-1/2}_{nn}} \int^{+1/2}_{-1/2} f^{\prime 2} (\tau ) \d \tau .
\end{equation}

So we get factors in the path integral
\begin{equation}\label{expM0}
\exp \left [ {i\over \varepsilon} (\Delta \gsth \Msth \Delta \gsth ) \right ].
\end{equation}

\noindent The form $(\Delta \gsth \Msth \Delta \gsth )$ reminds simplicial DeWitt supermetric on 3-geometries (see, e. g., Ref. \cite{HamWil}). Using (\ref{delta-x}) and
\begin{equation}
\det \Msth = - (\det \gsth )^{-4}/4
\end{equation}

\noindent we find
\begin{eqnarray}\label{expM}
& & \exp (i S_{\rm g \, sing}) = \exp \left [ {i\over \varepsilon} (\Delta \gsth \Msth \Delta \gsth ) \right ] \\& & \hspace{-10mm} = 2 (\pi\varepsilon)^3 (\det \gsth )^2 \delta^6 (\Delta \gsth ) \sim \Vsth^4 \delta^6 (\Dsth \ltwosth ). \nonumber
\end{eqnarray}

\noindent Here $\ltwosigd$ is the set of linklengths squared of the $d$-simplex $\sigd$. In particular, $\ltwoson$ is ${\rm (length)}^2$ of the edge $\son$. We should take product of these factors over $\sth$,
\begin{equation}\label{prodsth}
\prod_{\sth} \Vsth^4 \delta^6 (\Dsth \ltwosth).
\end{equation}

\noindent This contains $\delta$-functions of dependent arguments and thus $\delta$-functions of zero. These are connected with cycles around triangles $\stw$. Regularized \dfuns of zero are finite,
\begin{equation}\label{delta-sig3-0}
\dsth^6 (0) \equiv \delta^6 (\Dsth \ltwosth ) |_{\Dsth \ltwosth = 0} \sim \Vsth^{-4},
\end{equation}

\noindent see (\ref{expM}). Analogously for the face $\sigd \subset \sth$
\begin{equation}\label{delta-sigd-0}
\dsigd^{d(d+1)/2} (0) \equiv \delta^{d(d+1)/2} (\Dsigd \ltwosigd ) |_{\Dsigd \ltwosigd = 0} \sim \Vsigd^{-d-1}
\end{equation}

\noindent (considering $\Msth$ reduced to subspace spanned by $\sigd$). With the help of a routine combinatory consideration we can write (\ref{prodsth}) as
\begin{equation}\label{prodprime}
\prod_{\sth} \Vsth^4 \left ( \prod_{\son,\sth \supset \son}\!\!\!\!\!\!^{\prime} \delta (\Dsth \ltwoson) \right )\prod_{\stw} \dthstwnul \left( \prod_{\son} \dsonnul \right )^{-1},
\end{equation}

\noindent see Appendix. The primed product here means that redundant \dfuns are omitted (i. e. it is well-defined). By (\ref{delta-sigd-0}) this reads
\begin{equation}\label{res}
\prod_{\sth}\Vsth^4 \prod_{\stw}\Vstw^{-3} \prod_{\son}\Vson^2 \prod_{\son,\sth \supset \son}\!\!\!\!\!\!^{\prime} \delta (\Dsth \ltwoson).
\end{equation}

Besides $S_{\rm g \, sing}$, we have in $S_{\rm g}$ the finite at $\varepsilon \to 0$ part $S_{\rm g \, fin} = S_{\rm g} - S_{\rm g \, sing}$ which when appearing in the path integral exponential in the presence of the \dfun pre-exponent (\ref{res}) is Regge action. Just as $\Sind$ of Section \ref{1dim-example} considered as the sum over independent parts of the trajectories, the $S_{\rm g \, fin}$ can be taken as a sum of the contributions of a priori independent 4-simplices,
\begin{equation}
S_{\rm g \, fin} \! \propto \! \sumstw \varphi_{\stw} \Astw \! = \! \sumsfo \! \sumstwsubsfo \! \left ( \frac{2 \pi}{ \Nstw } - \alstwsfo \! \right ) \! \Astwsfo, ~~ \varphi_{\stw} \! = \! 2\pi - \!\!\! \sumsfosupstw \! \alstwsfo.
\end{equation}

\noindent Here $\alstwsfo$ is hyperdihedral angle of the 4-simplex $\sfo$ at the 2-face $\stw$, $\varphi_{\stw}$ is the defect angle at $\stw$, $\Nstw$ is the number of the 4-simplices meeting at $\stw$, $\Astwsfo$ is area of $\stw$ in the 4-simplex $\sfo$. It is taken into account that the value $\Astwsfo$ is independent of $\sfo \supset \stw$ in the presence of the \dfun pre-exponent (\ref{res}): $\Astwsfo \equiv \Astw$. 

Therefore it is natural that the corresponding to $S_{\rm g \, fin}$ path integral measure be the product of the independent path integral measures over the 4-simplices $\d \mu (\sfo)$, like $\dmuind$ of Section \ref{1dim-example} being the product of the independent measures over parts of the path. So the measure in simplicial gravity obtained from generalization of simplicial gravity ansatz to arbitrarily sharp metric jumps would read
\begin{equation}\label{dmug}
\d \mu_{\rm g} = \prod_{\sfo} \d \mu (\sfo) \prod_{\sth}\Vsth^4 \prod_{\stw}\Vstw^{-3} \prod_{\son}\Vson^2 \prod_{\son,\sth \supset \son}\!\!\!\!\!\!^{\prime} \delta (\Dsth \ltwoson).
\end{equation}

\noindent Without the last factor (\dfuns) in the formula (\ref{dmug}) we would have the set of independent 4-tetrahedra, while with this factor we get usual simplicial geometry and such effects as gravitons etc.

\section{Conclusion}

Thus, considering the arbitrarily sharp jumps of the simplicial metric between the 4-simplices leads to some rules for gluing separate independent measures on the 4-simplices into the one for the whole simplicial complex. Of course, here one else important and physically quite natural postulate is implied, namely, that the measure, corresponding to the action representable as a sum of independent terms, should be the product of independent measures. Thereby the path integral measure for simplicial gravity is related to the product of independent measures over 4-simplices.
We see that there is a correspondence of the discrete measure (obtained with the help of the physical assumptions just mentioned) with the continuum version. Namely, the latter is a product over points $x$ of some local measures, e. g. $\prod_x (- \det g)^{5/2} \prod_{\lambda \geq \mu} \d g_{\lambda \mu}$ (Misner measure \cite{Mis}). The product over 4-tetrahedrons for the discrete version looks quite expectable since the 4-tetrahedron in simplicial gravity is natural analog of the point in the continuum theory as elementary object where the metric could be defined.

The question may arise why can not we relate the path integral measure in the analogous way to the product of independent factors for any field different from gravity. Peculiar feature of gravity is that physical curvature distribution has 2-dimensional support while "virtual" curvature responsible for independence of neighboring 4-tetrahedra develops in the interior of 3-faces, that is, lives in another points. It would be quite unnatural to suggest that effect of these points of 3-faces depends on another points, those of boundary 2-faces (i. e. that $\varepsilon$ depends on defect angles). Contrary to that, physical non-gravity field such as electromagnetic one, is distributed everywhere over 4-volume, including points of 3-faces. Therefore corresponding $\varepsilon$ well may depend on characteristics of this field. The problem of specifying $\varepsilon$ and thus the measure would require additional assumptions in this case.

Mathematically exhaustive solution to the problem without operating with such objects as \dfuns of zero requires studying the limit of the product of the factors in the path integral even more general than (\ref{expM0}) (obtained by replacing $\varepsilon \to \varepsilon_{\sth}$),
\begin{equation}
\prod_{\sth} \exp \left [ {i\over \varepsilon_{\sth}} (\Delta \gsth \Msth \Delta \gsth ) \right ],
\end{equation}

\noindent at $\varepsilon_{\sth} \to 0$. The difficulty is, generally speaking, dependence of the result on the specific way of performing this limit, in particular, on the priority of tending $\varepsilon_{\sth} \to 0$ for different $\sth$s. On the other hand, this absence of strictly fixed answer just allows to put forward the conjecture that the method of performing this limit exists leading to the above result (\ref{dmug}).

In particular, the symmetry of the result obtained (w.r.t. the different simplices) is due to the above choice (\ref{varepsilon}) of dimensionless $\varepsilon$ (which absorbs $\Vsth$) as regularization parameter not depending on the geometrical characteristics of the different 3-faces $\sth$. Were it not so, above $\dsigd^{d(d+1)/2} (0)$, $d = 1,2$ would depend not only on $\sigd$ (\ref{delta-sigd-0}), but also on the choice of $\sth \supset \sigd$, and the considered symmetry w.r.t. the different simplices would not be achievable.

\section*{Acknowledgments}

The present work was supported in part by the Russian Foundation for Basic Research
through Grants No. 08-02-00960-a and No. 09-01-00142-a.

\section*{Appendix}

The product of \dfuns over all 3-simplices $\sth$ (usual tetrahedrons)
\begin{equation}\label{prod-delta}
\prod_{\sth} \delta^6 (\Dsth \ltwosth)
\end{equation}

\noindent contains $\delta$-functions of dependent arguments and thus $\delta$-functions of zero. These are connected with cycles around triangles $\stw$. Given a link $\son$ and a
triangle $\stw \supset \son$, the lengths squared of $\son$ in the 4-tetrahedra $\sfo_1 , \dots , \sfo_n \supset \stw$ enter the product (\ref{prod-delta}) through the product of $\delta$-functions of discontinuities of $\ltwoson$ on 3-faces $\sth_k = \sfo_k \cap \sfo_{k+1}$ between these 4-tetrahedra, $\Dsthk \ltwoson = \ltwosonsfok - \ltwosonsfokplon$, $k$ = 1,2, \dots, $n$ ($\sfo_{n+1} \equiv \sfo_1$). This product is not well-defined for it contains $\delta$-function of zero,
\begin{equation}
\prod^n_{k=1} \delta (\ltwosonsfok -\ltwosonsfokplon ) = \delta (0) \prod^n_{k=1}\! ^{\prime} \delta (\ltwosonsfok -\ltwosonsfokplon ).
\end{equation}

\noindent Here $\prod^{\prime}$ is well-defined product of $\delta$-functions obtained by omitting any one of the factors $\delta (\Dsthk \ltwoson)$. There are two else edges forming the given triangle $\stw$ so we get $\dthstwnul \equiv \delta^3 (\Dstw \gstw = 0)$ in (\ref{prod-delta}). Here subscript $\stw$ indicates that regularized $\delta^3 (\Dstw \gstw )$ at $\Dstw \gstw = 0$ depends on $\stw$, see (\ref{delta-sigd-0}). If, however, we take product of $\dthstwnul$ over all $\stw$ sharing the edge $\son$, this set of \dfuns of zero is larger than that contained in (\ref{prod-delta}). Indeed, let $N$ be number of the triangles $\stw \supset \son$. Let $\sth_k$, $k = 1,2, \dots , N-1$ be chain of $N-1$ tetrahedrons successively passing through these $\stw$. That is, if numbered accordingly, $\stw_k = \sth_{k-1} \cap \sth_k$, $k = 2,3, \dots , N-1$, and $\stw_1 \subset \sth_1$ and $\stw_N \subset \sth_{N-1}$ are ending triangles in this chain. Then the product of $\delta (\Dsth \ltwoson )$ over $\sth \supset \son$, $\sth \not\in \{\sth_k | k = 1,2, \dots, N-1\}$ can be well-defined in (\ref{prod-delta}). Only remaining $N-1$ \dfuns $\delta (\Dsthk \ltwoson)$, $k = 1,2, \dots, N-1$ acquire zero arguments, whereas product of $\dthstwnul$ over $\stw \supset \son$ contains just $N$ such functions. The result of throwing away extra factor $\dsonnul$ can be written as
\begin{equation}
(\dsonnul )^{-1}\prod_{\stw \supset \son} \dthstwnul.
\end{equation}

\noindent Considering this for all edges $\son$, we can write (\ref{prod-delta}) as
\begin{equation}
\prod_{\sth} \Vsth^4 \left ( \prod_{\son,\sth \supset \son}\!\!\!\!\!\!^{\prime} \delta (\Dsth \ltwoson) \right )\prod_{\stw} \dthstwnul \left( \prod_{\son} \dsonnul \right )^{-1}.
\end{equation}

\noindent The primed product here means that redundant \dfuns are omitted (i. e. it is well-defined).



\begin{thebibliography}{99}
\bibitem{Regge}
 Regge, T.: General relativity theory without coordinates. Nuovo Cimento {\bf 19}, 558 (1961).
\bibitem{Fri}
 Friedberg, R., Lee, T.D.: Derivation of Regge's action from Einstein's theory of general relativity. Nucl. Phys. B, {\bf 242}, 145 (1984).
\bibitem{Kha}
 Khatsymovsky, V.M.: Regge calculus from discontinuous metrics. Phys. Lett. B,
 {\bf 567} 288, (2003). arXiv:gr-qc/0304006 (2003).
\bibitem{WaiWil}
 Wainwright, C., Williams, R.M.: Area Regge calculus and discontinuous metrics. Class. Quantum Grav., {\bf 21}, 4865 (2004). arXiv:gr-qc/0405031 (2004).
\bibitem{HamWil}
 Hamber, H.W., Williams, R.M.:  Gauge Invariance in Simplicial Gravity. Nucl. Phys. B, {\bf 487}, 345 (1997). arXiv:hep-th/9607153 (1996).
\bibitem{Mis}
 Misner, C.W.:  Feynman quantization of general relativity. Rev. Mod. Phys., {\bf 29}, 497 (1957).
\end{thebibliography}
\end{document}